\documentclass[onecolumn,showpacs,preprintnumbers,showkeys,amsmath,amssymb]{revtex4}
\usepackage{epsfig}
\usepackage{dcolumn}
\usepackage{bm}

\begin{document}

\title{Monte Carlo studies of the square Ising model with next-nearest-neighbor interactions}

\author{A. Malakis}
\altaffiliation[]{Corresponding author: amalakis@cc.uoa.gr}
\author{P. Kalozoumis}
\author{N. Tyraskis}
\affiliation{Department of Physics, Section of Solid State
Physics, University of Athens, Panepistimiopolis, GR 15784
Zografos, Athens, Greece}

\date{\today}

\begin{abstract}
We apply a new entropic scheme to study the critical behavior of
the square-lattice Ising model with nearest- and
next-nearest-neighbor antiferromagnetic interactions. Estimates of
the present scheme are compared with those of the Metropolis
algorithm. We consider interactions in the range where
superantiferromagnetic (SAF) order appears at low temperatures. A
recent prediction of a first-order transition along a certain
range (0.5-1.2) of the interaction ratio $(R=J_{nnn}/J_{nn})$ is
examined by generating accurate data for large lattices at a
particular value of the ratio $(R=1)$. Our study does not support
a first-order transition and a convincing finite-size scaling
analysis of the model is presented, yielding accurate estimates
for all critical exponents for $R=1$ . The magnetic exponents are
found to obey  ``weak universality'' in accordance with a previous
conjecture.
\end{abstract}

 \maketitle

\section{Introduction}
Spin models with competing interactions are of great theoretical
and experimental interest and have been for long time the subject
of many investigations. Such interactions may give rise to
complicated spatial orderings and produce complex and rich
critical behavior~\cite{selke}. The transitions between ordered
and disordered phases may be continuous or of first order with a
tricritical point between them, but no general classification
exists, connecting the symmetry of spin structures and the range
of interaction with the expected critical behavior. The Ising
square lattice with nearest-neighbor coupling (nn) is an exactly
soluble model~\cite{onsager} and almost all its properties are
well known. With the addition of next-nearest-neighbor (nnn)
interactions the problem is no longer exactly soluble and several
approximate methods have been applied to attack this more general
problem and to understand the effect of adding the nnn-coupling on
the critical behavior of the
system~\cite{nightingale,swedensen,binder,landau,oitmae,landbind,minami}.
Of particular interest is the case of competing interactions,
where the ground state is an arrangement with
superantiferromagnetic (SAF) order in which ferromagnetic rows
(columns) alternate with opposite oriented spins. The $T=0$ phase
diagram is well known~\cite{binder,landau,oitmae} and the
SAF-order can be obtained in both cases of a ferromagnetic or an
antiferromagnetic nn-coupling. The system, in zero-field, is
governed by the Hamiltonian:
\begin{equation}
\mathcal{H}=J_{nn}\sum_{<i,j>}S_{i}S_{j}+J_{nnn}\sum_{(i,j)}S_{i}S_{j}
\end{equation}
where here both nearest-neighbor $(J_{nn} )$ and next
nearest-neighbor $(J_{nnn})$ interactions will be assumed to be
positive (antiferromagnetic) and the system as is well
known~\cite{nightingale,swedensen,binder,landau,oitmae,landbind,minami}
develops at low temperatures superantiferromagnetic order for
$R>0.5$.

Several previous studies have suggested that the above system may
possess ``anomalous'' exponents, and a non-universal critical
behavior with exponents depending on the coupling ratio
$R=J_{nnn}/J_{nn}$ has been the commonly accepted scenario for
many
years~\cite{nightingale,swedensen,binder,landau,oitmae,landbind,minami}.
However, recently the interest on the subject has been renewed and
some attempts to re-examine the behavior of this model have taken
place. In several papers Lopez
et.al~\cite{lopez93,lopez94,lopez99} have used the cluster
variation method (CVM) to study this model and have concluded that
the system undergoes a first-order transition for a particular
range of the coupling ratio (0.5-1.2). Thus, a different scenario
predicting first order transitions between ordered and disordered
phases, followed by continuous transitions outside the first-order
region has been proposed~\cite{lopez93,lopez94,lopez99}. It
appears that this scenario has been further supported by the study
of Buzano and Pretti~\cite{buzano}. These authors studied the same
model with an additional 4-body coupling using again the CVM and
concluded that according to this method a first-order behavior is
expected for a very large part of the parameter space reproducing
also the results of~\cite{lopez93,lopez94,lopez99}. However, they
also considered the limiting case $(J_{nn}=0)$, where the exact
solution of Baxter model~\cite{baxter} applies, observing again a
large part of the parameter space in which the CVM predicts
first-order behavior. Thus, the CVM fails to predict the true
second order critical behavior for the Baxter model and, this is
of course, an obvious reason for suspecting the CVM. It is quite
possible that the CVM, in any finite-order approximation, could
produce misleading ``mean-field'' behavior not unlikely with other
variational methods~\cite{malakis}. Yet, the relevant questions
are very important for a better understanding of the critical
behavior of systems with competing interactions. Notwithstanding
that, a recent Monte Carlo simulation~\cite{rastelli} of a quite
``similar model'' has provided strong evidence of first order
behavior. This last model includes nn- and nnn-antiferromagnetic
couplings but now the model is defined on the triangular lattice.
Having to deal with the same hamiltonian and a similar ground
state arrangement (SAF-arrangement) one should naively expect the
same critical behavior for the two models, assuming a simple
universality in which the underlying lattice is immaterial in 2D.
In the light of the above controversial and at the moment
unsettled situation, it is of interest to follow again the
traditional finite-size scaling analysis to re-examine the above
prediction. As a first attempt, we study here the square model for
a particular value of the coupling ratio, $R=1$, included in the
above range, and using a new complementary and efficient numerical
method we generate accurate finite-size data for quite large
lattices. The rest of the paper is organized as follows. In the
next section we outline the entropic sampling technique
implemented here to generate numerical estimates accurate in the
critical region for the case $R=1$. In section 3 we explain why
our numerical data are not supporting a first-order transition and
then we present a finite-size analysis, yielding accurate
estimates for the critical exponents. These are compared with
estimates existing in literature and an old conjecture of K.
Binder and D. P. Landau~\cite{binder} assuming ``weak
universality'' is reinforced. Our conclusions are summarized in
section 4.

\section{The CrMES Wang-Landau entropic sampling scheme}
\label{section2}

Traditional Monte Carlo sampling methods have been used for many
years in the study of critical phenomena~\cite{landbindbook} and
the first Monte Carlo approaches to the present model were indeed
carried out by the Metropolis
method~\cite{swedensen,binder,landau}. On the other hand, general
flat histogram  methods~\cite{landbindbook,wang,schulz,oliveira96}
are alternatives to importance sampling and are expected to be
much more efficient for studying a complex system. A simple and
efficient entropic implementation of the Wang-Landau(WL) method
has been presented recently by the present authors. This method is
based on a systematic restriction of the energy space as we
increase the lattice size. The random walk of this entropic
simulation takes place only in the appropriately restricted energy
space and this restriction produces an immense speed up of all
popular algorithms calculating the density of energy states (DOS)
of a statistical system~\cite{malakis04,malakis05}. For the
temperature range of interest, that is the range around a critical
point, this scheme determines all finite-size anomalies using an
one-run entropic strategy employing what we have called the
``critical minimum energy subspace Wang-Landau (CrMES WL)-entropic
sampling method''. In this approach all (thermal and magnetic)
properties are obtained by using the high-levels (that is the
levels where the detailed-balanced condition is quite well obeyed)
of the Wang-Landau random walk process to determine appropriate
microcanonical estimators. The method is efficiently combined with
the N-fold way [19] in order to improve statistical reliability
but also to produce broad-histogram (BH) estimators for an
additional calculation of the density of states (DOS).

The approximation of canonical averages, in a temperature range of
interest, is as follows:
\begin{eqnarray}
\langle Q\rangle=\frac{\sum_{E}\left< Q\right>_{E}G(E)e^{-\beta
E}}{\sum_{E}G(E)e^{-\beta E}}\cong
\frac{\sum_{E\in(E_{1},E_{2})}\langle
Q\rangle_{E,WL}\widetilde{G}(E)e^{-\beta
E}}{\sum_{E\in(E_{1},E_{2})}\widetilde{G}(E)e^{-\beta E}}
\end{eqnarray}

The restricted energy subspace $(E_{1},E_{2})$  is carefully
chosen to cover the temperature range of interest without
introducing observable errors. The microcanonical averages
$\langle Q\rangle_{E}$ are determined from the
$H_{WL}(E,Q)$-histograms, which are obtained during the
high-levels of the WL-process:
\begin{equation}
\langle Q\rangle_{E}\cong\langle Q\rangle_{E,WL}\equiv
\sum_{Q}Q\frac{H_{WL}(E,Q)}{H_{WL}(E)},\;
H_{WL}(E)=\sum_{Q}H_{WL}(E,Q)
\end{equation}
and the summations run over all values generated in the restricted
energy subspace $(E_{1},E_{2})$. Finally, the approximate density
of states  used in (2) is obtained from the DOS generated after
the last WL-iteration $(\widetilde{G}(E)=G_{WL}(E))$ or from the
broad-histogram (BH) approximation $(\widetilde{G}(E)=G_{BH}(E))$.
This last approximation may be easily calculated from appropriate
microcanonical averages corresponding also to the high-levels of
the WL-process, since in the N-fold implementation of the
WL-process the necessary histograms for the application of the
broad-histogram method~\cite{oliveira96}
 are also known. As mentioned above, the updating of appropriate histograms
(Q may be any power of the order parameter or some other quantity)
is carried out only in the high-levels of the WL-process. In these
stages, the incomplete detailed-balance condition has not
significant effect on the microcanonical estimators constructed
from the cumulative histograms as shown in~\cite{malakis05}. Thus
we have used only the WL-iterations:  $j=12-24$ for lattices up to
$L=100$ and the WL-iterations: $j=16-26$ for larger lattices. The
initial modification factor of the WL-process is taken to be
$f_{1}=e=2.718...$ and, as usual, we follow the rule
$f_{j+1}=\sqrt{f_{j}}$ and a $5\%$ flatness
criterion~\cite{malakis04,malakis05}. The rest of the details and
the N-fold implementation can be found
in~\cite{landbindbook,wang,schulz,oliveira96,malakis04,malakis05}.

In the present implementation of the CrMES method we restrict the
total energy range $(E_{min},E_{max})$ to the minimum
energy-subspace producing an accurate estimation for all
finite-size anomalies. This restriction may be defined by
requesting a specified accuracy on a diverging specific heat (or
on a diverging susceptibility) as shown in \cite{malakis05}.
Alternatively, the energy density function may be used in a
simpler way to restrict the energy space. Thus, if  $\tilde{E}$ is
the value maximizing the probability density, at some
pseudocritical temperature $(T_{L}^{*})$, the end-points
$(\widetilde{E}_{\pm})$ of the energy critical subspaces (CrMES)
may be located by the condition:
\begin{equation}
\widetilde{E}_{\pm}:
\frac{P_{\widetilde{E}_{\pm}}(T_{L}^{*})}{P_{\widetilde{E}}(T_{L}^{*})}\leq
r
\end{equation}
where  r is chosen to be a small number, independent of the
lattice size. Both  $r=10^{-4}$ and $r=10^{-6}$ have been used in
the present study to estimate the relevant extensions of critical
subspaces as discussed bellow. The resulting finite-size
extensions of critical energy subspaces, denoted by
$(\Delta\widetilde{E})$ , obey the ``specific-heat'' scaling
law~\cite{malakis04}:
\begin{equation}
 \Psi\equiv
\frac{(\Delta\widetilde{E})^{2}}{L^{d}}\approx
L^{\frac{\alpha}{\nu}}
\end{equation}
Since the extensions of these energy subspaces satisfy very well
the above scaling law, relation (5) is a new route for estimating
the critical exponent $\alpha/\nu$~\cite{malakis04,malakis05}.
Furthermore, following a similar procedure we may also estimate
the critical exponent $\gamma/\nu$, as already shown
in~\cite{malakis05}. Again, let $\widetilde{M}$ be the value
maximizing the order parameter density at some pseudocritical
temperature, for instance at the susceptibility pseudocritical
temperature. The end-points $(\widetilde{M}_{\pm})$ of the
magnetic critical subspaces (CrMMS) are located by the condition:
\begin{equation}
\widetilde{M}_{\pm}:
\frac{P_{\widetilde{M}_{\pm}}(T_{L}^{*})}{P_{\widetilde{M}}(T_{L}^{*})}\leq
r
\end{equation}
and  the corresponding finite-size extensions  of critical
magnetic subspaces obey close to a critical point, the
``susceptibility'' scaling law~\cite{malakis05} :
\begin{equation}
\Xi\equiv \frac{(\Delta\widetilde{M})^{2}}{L^{d}}\approx
L^{\frac{\gamma}{\nu}}
\end{equation}
Equations (equations (5) and (7)) provide additional routes for
estimating the involved critical exponents.

\section{Numerical evidence. Finite - size scaling analysis.}
\label{section3}

Let us first present an illustrative graph showing that the
above-described entropic scheme yields accurate data for the
application of finite-size scaling. We have used two different
definitions for the order parameter. With the help of four
sublattices of the SAF-ordering one may define a two-component
order parameter and finally use its root-mean-square (rms) as done
in~\cite{binder}.
\begin{eqnarray}
&M_{SAF}^{(1)}=\left\{M_{1}+M_{2}-(M_{3}+M_{4})\right\}/4,\;\;\;\;
M_{SAF}^{(2)}=\left\{M_{1}+M_{4}-(M_{2}+M_{3})\right\}/4\nonumber\\
&M_{SAF}^{(rms)}=\sqrt{\left(M_{SAF}^{(1)}\right)^{2}+\left(M_{SAF}^{(2)}\right)^{2}}
\end{eqnarray}
We have used this rms order parameter and also, as an alternative,
the sum of the absolute values of the four sublattice
magnetizations $(M_{SAF}=\sum_{i=1}^{4}\left|M_{i}\right|/4)$. The
resulting behavior is very similar and the finite size extensions
of the resulting CrMMS completely coincide supporting the identity
of the two representations for the present system. Therefore, for
large lattices only the second order parameter was used. For a
particular temperature, $T=2.082$ , close to the critical
temperature, we have calculated using long runs of the Metropolis
algorithm several thermodynamic properties of the system for $R=1$
and we have found good agreement with the corresponding estimates
obtained via the entropic scheme described in the previous
section. Figure 1 provides a comparison test between the
Metropolis algorithm and entropic scheme. Let us consider the
estimates of the Metropolis simulation for the susceptibility at
the above mentioned temperature as being the exact values. Then
figure 1 presents the variation of the ``relative errors'' of the
entropic scheme with lattice size and compare these with the
statistical errors of the Metropolis simulation. From this figure
it can be seen that the CrMES-WL
 entropic scheme provides estimates in the
expected range. We conclude that similar accuracy should be
expected for the temperature-range covered by the restricted
energy subspaces.

Next, we search (using the approximate DOS: $\widetilde{G}(E)$)
for a double peak in the energy probability density which should
be expected if the system undergoes a first-order transition as
predicted by CVM. For $R=1$, using small temperature steps close
to the specific heat pseudocritical temperatures, we did not
observe such double peaks. In contrast when we used our algorithm
to generate the appropriate DOS of the triangular (SAF) model
considered by Rasteli et. al ~\cite{rastelli}, the presence of the
energy double-peaks reported by these authors was very clear. The
finite-size behavior of the fourth-order cumulant (of the order
parameter) is indicative for the order of the
transition~\cite{binderlett,shanho}. We used this test in order to
observe the behavior for the square lattice model and to examine
the prediction for a first-order transition reported by Lopez
\emph{et al.}~\cite{lopez93,lopez94,lopez99}. Comparing by this
test the two models we had the opportunity to observe that the
difference in the cumulant-behavior between them was again
profound. For the triangular model the behavior was in very good
agreement with that reported in~\cite{rastelli} indicating a
first-order transition, while for the present square model a
behavior characteristic of a second order critical point was
observed~\cite{binderlett}. It appears that the first-order
prediction of the CVM is false at least for the case $R=1$. The
tricritical point, if it exists, could be in a lower temperature
and the upper bound (1.2) given by Lopez \emph{et
al.}~\cite{lopez93,lopez94,lopez99} could be an overestimation of
the CVM.

Figure 2 presents the order-parameter cumulant behavior of the
square lattice model for several lattice sizes close to the
critical temperature. From this graph and by using the crossing
method~\cite{binderlett} we have estimated the critical
temperature. Taking the average of all the crossing temperatures
included in figure 2 we find: $T_{c}=2.0823(17)$. Including in
this averaging the smaller sizes $L=30-60$ the estimate is:
$T_{c}=2.0821(13)$. Thus, $T_{c}=2.0823(17)$ seems quite safe and
it is also in good agreement with the estimates obtained by
fitting the specific heat, susceptibility and energy-cumulant
pseudocritical temperatures to a power law behavior with a
correction term:
\begin{equation}
T_{L}=T_{c}+\alpha L^{-\lambda}(1+\frac{b}{L})
\end{equation}
The above-mentioned fits, not shown for brevity, yielded
respectively: $(T_{c}=2.0825(5), \lambda=1.20(4))$,
$(T_{c}=2.0828(8), \lambda=1.197(50))$ and  $(T_{c}=2.0818(8),
\lambda=1.158(60))$

Let us now try to estimate the magnetic critical exponent
$\gamma/\nu$. We employ the commonly used route of fitting the
values of the susceptibility peaks and/or its values at the
critical temperature $(T_{c}=2.0823)$ to observe the scaling
exponent. Figure 3 presents three scaling fitting attempts
assuming a simple power law. The peaks of the susceptibility yield
an estimate of the order 1.79, while the susceptibility values at
the estimated critical temperature provide an estimate of the
order 1.71. The exponent appears to acquire a value in this wide
range $(\gamma/\nu=1.71-1.79)$ if we vary the lattice sizes fitted
and/or if we add corrections terms to the simple power law. Note
that the estimated in~\cite{binder} range is $1.71\pm 0.15$. The
middle solid line in figure 3 shows that the average of the
susceptibility in the two temperatures ($T_{c}$ and $T
_{L}^{*}(\chi)$) gives an estimate very close to the 2D-Ising
value $(\gamma/\nu=1.75)$. These observations seem to favor the
original view of~\cite{binder} that the system may obey a kind of
``weak universality''~\cite{suzuki}. According to this hypothesis
the reduced exponent $\gamma/\nu$  will have the 2D-Ising value
independent of R. To further check this conjecture let us analyze
the scaling behavior of the finite-size extensions
$(\Delta\widetilde{M})$ in equation (7). Figure 4 shows the
behavior of these scaled extensions in the two temperatures
$T_{c}$ and $T _{L}^{*}(\chi)$) together with the behavior of
their average. Fitting the numerical data to a law of the form:
\begin{equation}
y=\alpha L^{w}(1+\frac{b}{L})
\end{equation}
we obtain very good fits and the estimates of $\gamma/\nu$  are
for the three curves in the range $w=1.75 \pm 0.01$. Thus, the
scenario of weak universality of $\gamma/\nu=1.75$ is greatly
reinforced.

Frequently, the scaling of the specific heat is a difficult task
but for the present model we have discovered that the scaling
behavior of the average of our estimates in the two temperatures
$(T_{c}$ and $T _{L}^{*}(C))$
 is quite stable as one varies the
lattice size. Figure 5 shows details of the fit for the
``averaged''  specific heat values and figures 6 and 7 show the
analysis of the scaled extensions $(\Psi(\Delta\widetilde{E}))$
appearing in equation (5). Comparing the last two fits we find a
unique case of stability and we confidently estimate:
\begin{equation}
\frac{\alpha}{\nu}=0.412\pm0.005
\end{equation}
The above findings (estimate (11) and the weak universality
$\gamma/\nu=1.75$) may now be used to determine a consistent
scheme for all critical exponents. Assuming hyperscaling, the
correlation length exponent is estimated as $\nu=0.8292(24)$  and
this value is consistent $(\frac{1}{\nu}=\lambda)$ with the
estimates of the shift exponent found from the pseudocritical
temperatures. The exponent $\beta/\nu$ should therefore be 0.125.
A power law fit of our estimates of the order parameter at the
critical temperature are in a good agreement with this value. The
proposed set of exponents satisfies all scaling laws for a second
order transition as can be easily verified.

\section{Concluding remarks}
\label{section4}

In this paper we have considered the square-lattice Ising model
with nearest- and next-nearest-neighbor antiferromagnetic
interactions and found that the prediction of the cluster
variation method of a first-order transition is not supported by
the finite size behavior of the system $(R=1)$. The original
scenario~\cite{nightingale,swedensen,binder,landau,oitmae,landbind,minami}
of a non-universal critical behavior with exponents depending on
the coupling ratio has been strongly reinforced by our numerical
study and ``weak universality''~\cite{binder,suzuki} seems to be
well obeyed. The idea of using scaled extensions of dominant
critical subspaces to estimate the thermal exponent $\alpha/\nu$
and the magnetic exponent $\gamma/\nu$ seems to supply a quite
accurate route for exponent estimation. In the present case the
accurate estimation by this route of the thermal exponent is
almost unique. Furthermore, the successful application of the
scaling law (7) has been a helpful guidance in observing the
``weak universality'' of the magnetic critical exponents. The
entropic sampling scheme applied in this study provides a better
evaluation of the tail behavior of the critical distributions and
this seems to be of importance for the accurate estimation of the
finite-size extensions of the dominant subspaces used for the
exponent estimation.

\section*{ACKNOWLEDGEMENTS}

This research was supported by the Special Account for Research
Grants of the University of Athens under Grant Nos. 70/4/4071.

{}

\begin{figure}
\includegraphics*[width=10 cm]{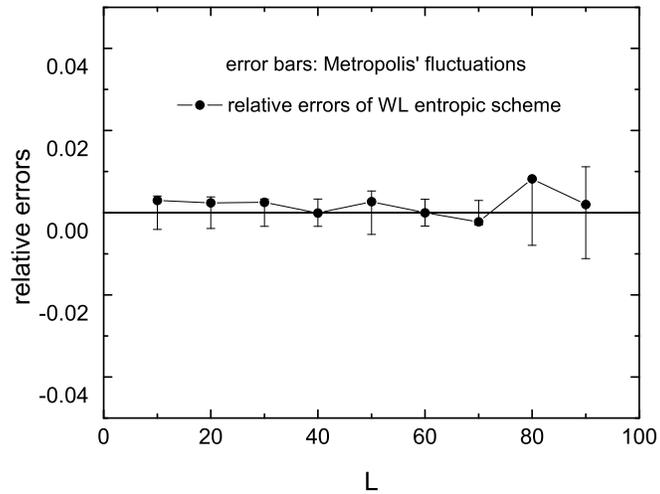}
\caption{\label{fig1}Relative deviations of the CrMES entropic
scheme used in this paper with respect to the Metropolis
algorithm. The relative errors of the critical susceptibility are
presented for $L=10-90$ and are compared to the statistical errors
(indicated as error bars) of the corresponding Metropolis runs.}
\end{figure}

\begin{figure}
\includegraphics*[width=10 cm]{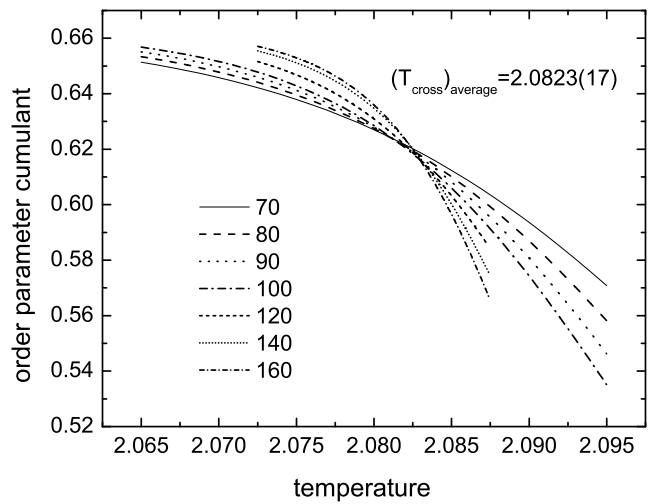}
\caption{\label{fig2}Estimation of the critical temperature
$(R=1)$. $T_{c}$ is obtained as the average of the crossing
points. The reduced fourth-order cumulant has been used.}
\end{figure}

\begin{figure}
\includegraphics*[width=10 cm]{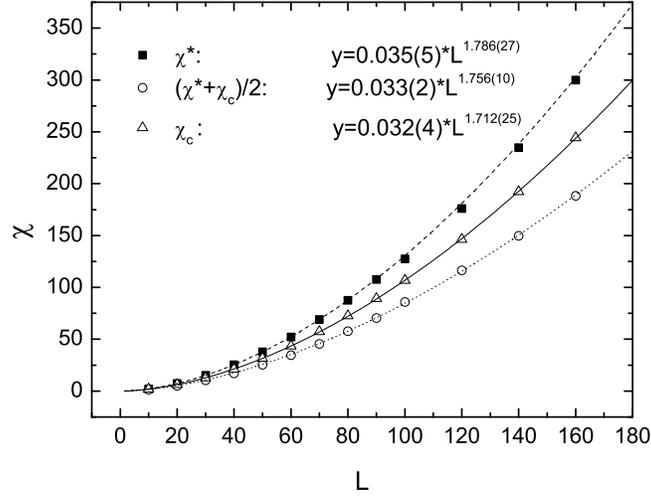}
\caption{\label{fig3}Finite-size behavior of the susceptibility at
the critical and its pseudocritical temperature. Fitting
parameters to a simple power law are presented for the above
estimates as well as for their average. Note, that only their
average gives an exponent value very close to the 2D Ising value.}
\end{figure}

\begin{figure}
\includegraphics*[width=10 cm]{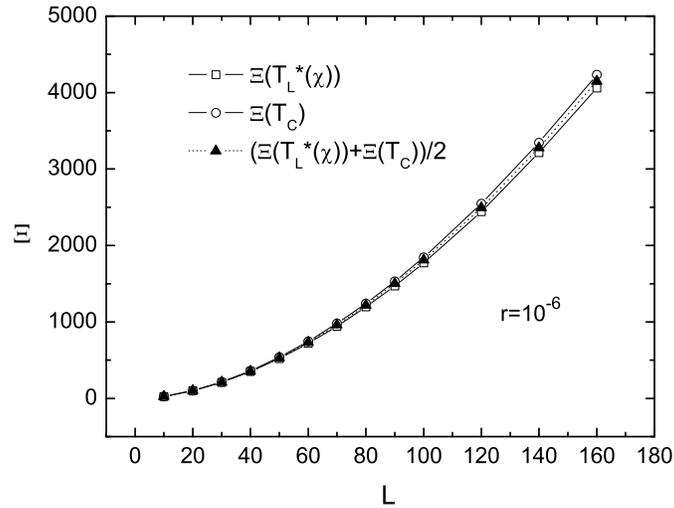}
\caption{\label{fig4}Finite-size behavior of the scaled extensions
$\Xi(\Delta\widetilde{M})$ at the critical and susceptibility's
pseudocritical temperature. Their average is also shown. Fitting
Eq.~(10) to these data we have obtained very good estimates of the
critical exponent $\gamma/\nu$: 1.7545(27), 1.7536(45), 1.7541(28)
respectively. These three estimates  fall into a narrow range
close to the 2D Ising value contrary to the wide range observed in
figure 3.}
\end{figure}

\begin{figure}
\includegraphics*[width=10 cm]{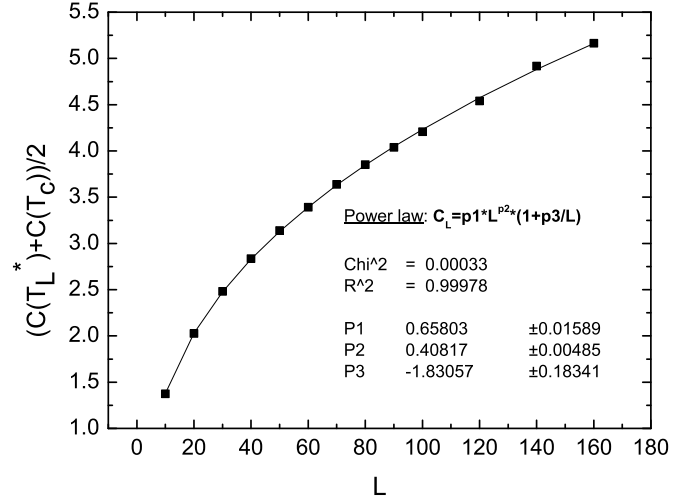}
\caption{\label{fig5}Finite-size behavior of the specific heat's
average at the critical and its pseudocritical temperature.
Fitting parameters to the power law illustrated on the graph, are
presented as well.}
\end{figure}

\begin{figure}
\includegraphics*[width=10 cm]{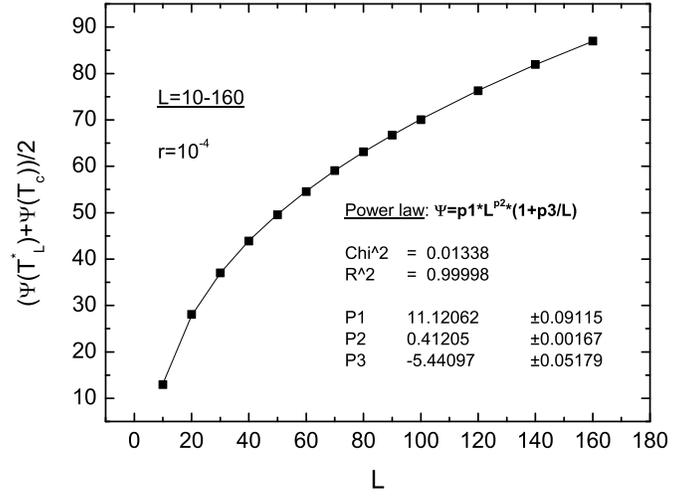}
\caption{\label{fig6}Finite-size behavior of the average of scaled
extensions $\Psi(\Delta\widetilde{E})$ (Eq.~(5)) at the critical
and and specific heat's pseudocritical temperature. Fitting
parameters to the power law illustrated on the graph, are
presented as well. The results are obtained through fitting all
the available data}
\end{figure}

\begin{figure}
\includegraphics*[width=10 cm]{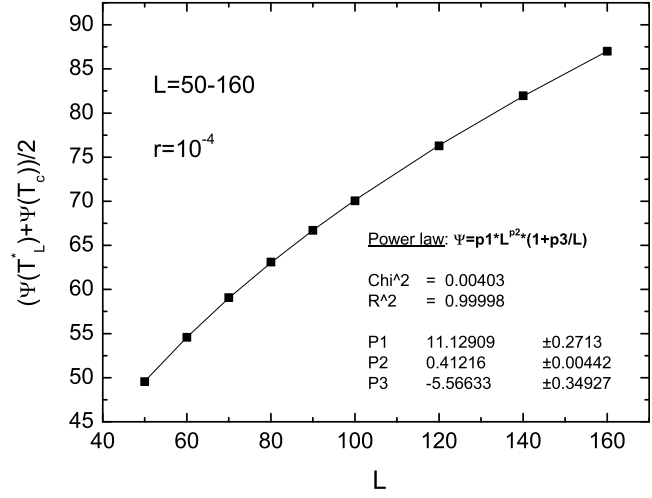}
\caption{\label{fig7}Finite-size behavior of the average of scaled
extensions $\Psi(\Delta\widetilde{E})$ at the critical and and
specific heat's pseudocritical temperature. The results are
obtained through fitting data from $L=50$ to $160$ in order to
check the stability of the fit.}
\end{figure}

\end{document}